\documentstyle[12pt]{article}
\begin{document}

\begin{titlepage}

\hfill{December 1996}


\hfill{UM-P-96/102}

\hfill{RCHEP-96/17}

\vskip 1.9 cm

\centerline{{\large \bf 
The exact parity symmetric model and big bang nucleosynthesis}}
\vskip 0.2cm

\vskip 1.3 cm
\centerline{R. Foot and R. R. Volkas}

\vskip 1.0 cm
\noindent
\centerline{{\it Research Centre for High Energy Physics,}}
\centerline{{\it School of Physics, University of Melbourne,}}
\centerline{{\it Parkville, 3052 Australia. }}

\vskip 1.0cm

\centerline{Abstract}
\vskip 0.7cm
\noindent
The assumption of exact, unbroken parity symmetry leads 
directly to a simple predictive resolution of the atmospheric
and solar neutrino puzzles. This is because the existence of 
this symmetry implies the existence of a set of
mirror neutrinos which must mix maximally with the known
neutrinos if neutrinos have mass.
The maximal mixing of the electron neutrino with the mirror
electron neutrino with $3 \times 10^{-10} \ eV^2 \stackrel{<}{\sim} 
|\delta m^2| \stackrel{<}{\sim} 10^{-3} \ eV^2$
leads to a predicted reduction of the solar neutrino flux 
by a factor of $2$, which is in quite good agreement with the experiments. 
The maximal mixing of the muon neutrino with the mirror muon neutrino
with $|\delta m^2| \simeq 10^{-2}\ eV^2$ also solves the
atmospheric neutrino puzzle.  We show that there is a significant 
range of parameters where these solutions are {\it not} in 
conflict with standard Big Bang Nucleosynthesis 
when the creation of lepton asymmetry due to neutrino 
oscillations is taken into account. 

\end{titlepage}

It has been known for a long time\cite{ly} but not
widely appreciated that it is possible to build a 
phenomenologically consistent gauge model which
has a parity symmetry which need not be broken at all.
In order to achieve {\it unbroken} parity 
symmetry it is necessary to double the number of fermions and the 
gauge symmetry.  However, while the number of particles is doubled the
number of parameters is not significantly increased (only two
additional parameters in the minimal model with massless
neutrinos)\cite{flv}. 
If the neutrinos in the exact parity symmetric model have mass, 
and if mass mixing between ordinary and mirror neutrinos exists, 
then the mass eigenstate neutrinos must also be parity eigenstates. 
Because parity transformations simply
interchange ordinary and mirror neutrinos, 
the mass eigenstate fields will be maximal combinations of
ordinary and mirror weak eigenstates\cite{flv}.
This result holds independently of the details of the origin
of the neutrino masses.  

The mirror neutrinos are essentially sterile
as far as ordinary interactions are concerned.
However unlike totally sterile neutrinos (such as right-handed
neutrino gauge singlets) mirror neutrinos interact 
amongst themselves and with mirror particles with
interactions of the same form and strength as ordinary
neutrinos interact with ordinary particles.
One advantage of mirror neutrinos over conventional sterile neutrinos
is that the mirror gauge symmetry provides an excellent reason
as to why they are not very heavy\cite{flv,bzm}.
If the parity symmetry connecting the ordinary and mirror worlds
is unbroken, then the mirror neutrinos are set by the same
scale as ordinary neutrinos\cite{flv}.

We will denote the three mirror neutrinos by 
\begin{equation}
\nu'_e, \ \nu'_{\mu}, \ \nu'_{\tau}.
\end{equation}
With small intergenerational mixing it follows 
from the unbroken parity symmetry of the model
that $\nu_e$ and $\nu'_e$ will be approximately
maximal mixtures of two mass eigenstates $\nu_1$ and
$\nu_2$.  Similarly, $\nu_{\mu}$ and $\nu'_{\mu}$ will each
be approximately maximal mixtures of two mass eigenstates
as will $\nu_{\tau}$ and $\nu'_{\tau}$. We will denote the 
$\delta m^2$ describing these maximal oscillations
by $\delta m^2_{ee'}, \ \delta m^2_{\mu \mu'}$ and $\delta 
m^2_{\tau \tau'}$ respectively.

The maximal mixing of $\nu_e$ and $\nu'_e$ will solve
the solar neutrino problem for the large range of 
parameters \cite{fv,pc,fn1}
\begin{equation}
3 \times 10^{-10}\ eV^2 \stackrel{<}{\sim}\ |\delta m^2_{ee'}| 
\stackrel{<}{\sim}\ 10^{-3} \ eV^2.
\label{solar}
\end{equation}
The maximal mixing of the electron neutrino with the mirror
electron neutrino with $\delta m^2_{ee'}$ in the above range
leads to a predicted reduction of the 
solar neutrino flux by a factor of $2$ 
which is in quite good agreement with the experiments. (See Ref.
\cite{fv} for a detailed comparison of the predictions of the
exact parity model with the solar neutrino data\cite{sn}). 

The deficit of atmospheric 
muon neutrinos can be explained if there are $\nu_{\mu}-\nu'_{\mu}$
oscillations with $\sin^2 2\theta_0 \stackrel{>}{\sim} 0.5$
and $ 10^{-3} \ eV^2 \stackrel{<}{\sim} |\delta m^2_{\mu \mu'}|
\stackrel{<}{\sim} 10^{-1}\ eV^2$ \cite{ana, barpak}.  
The best fit\cite{ana} occurs for 
$\sin^2 2\theta_0 \simeq 1$ and  
\begin{equation}
|\delta m^2_{\mu \mu'}| \simeq 10^{-2} 
\ eV^2. \label{atmos} \end{equation}
The exact parity symmetric model is also compatible with
the LSND signal\cite{lsnd,flv}.

A potential problem with any model that has additional light
degrees of freedom is that these extra states can contribute to
the energy density of the early Universe and spoil the 
reasonably successful Big Bang Nucleosynthesis (BBN) predictions.
This presents a problem for the exact parity symmetric model
because it can potentially lead to a doubling of the energy density at 
the time of nucleosynthesis (which is equivalent to about 6 additional
neutrinos). However, it is plausible that an initial
macroscopic asymmetry between ordinary and mirror matter might exist, as
can be arranged through the inflationary scenario proposed in 
Ref.\cite{Kolb} (for example). Even if ordinary matter dominates mirror
matter immediately after the Big Bang, the oscillations between the 
ordinary and mirror neutrinos might be expected to 
bring the mirror sector into equilibrium with the ordinary 
particles\cite{bb}. For maximally mixed ordinary - mirror neutrinos, the 
following BBN bounds have been obtained\cite{B} {\it assuming 
that the lepton number asymmetry could be neglected}:
\begin{equation}
|\delta m^2_{ee'}| \stackrel{<}{\sim} 10^{-8} \ eV^2,\
|\delta m^2_{\mu \mu'}|, \ |\delta m^2_{\tau \tau'}|  
\stackrel{<}{\sim} 10^{-6}\ eV^2.
\label{B}
\end{equation}
With the parameter choices Eq.(\ref{solar}, \ref{atmos}), there 
is a potential conflict with the naive BBN bounds Eq.(\ref{B}).
However, these bounds do not hold if there is an appreciable 
lepton asymmetry in the early Universe for 
temperatures between $1-30$ MeV\cite{fv1}.
Remarkably, it turns out that ordinary-sterile neutrino
oscillations can by themselves create lepton number \cite{ftv, shi, fv2}.
Recently, we have shown\cite{fv2} that the lepton number
generated by ordinary - sterile neutrino oscillations can
allow the bounds in Eq.(\ref{B}) to be evaded by many orders of
magnitude.  Indeed, the bounds can be relaxed sufficiently so that
the ordinary - sterile neutrino oscillation
solutions to the solar and atmospheric neutrino anomalies
do not significantly modify BBN.

The purpose of this paper is to study the special case where the
sterile neutrinos are mirror neutrinos\cite{flv}.
The mirror neutrinos are essentially sterile
when probed by ordinary matter, however they do have 
significant self interactions. There are two main effects of 
the self interactions in the early Universe.  First, the
effective potential governing ordinary - mirror
neutrino oscillations will gain a contribution from the
interactions of the mirror neutrino with the background.
Second, the mirror weak interactions can bring the mirror
neutrinos into equilibrium with the other mirror particles.
This effect is quite important because it will significantly
modify the momentum distribution and the number density of mirror neutrinos
compared with the case of sterile neutrinos.
In Ref.\cite{fv2}, we showed in detail how lepton number creation
can evade the naive bounds of Eq.(\ref{B}) and we determined the
required parameter space for strictly sterile neutrinos. In the
present work we consider the case of mirror neutrinos. For this case
it turns out that the effect of the mirror neutrino self interactions 
is to significantly enlarge the allowed region of parameter space compared 
to the case of strictly sterile neutrinos. 

There are many independent $\delta m^2, \ \sin^2 2\theta_0$ parameters.
We will need to make some assumptions otherwise we cannot say anything
definite.  We will assume that intergeneration mixing is small and that 
\begin{equation}
m_{\nu_e},\ m_{\nu'_e} < 
m_{\nu_{\mu}}, m_{\nu'_{\mu}} < m_{\nu_{\tau}},\
 m_{\nu'_{\tau}}. 
\label{hierarchy}
\end{equation}
The assumption of small mixing between the generations is quite natural
in our opinion in view of the situation with quarks. This assumption
is also supported by the LSND experiment\cite{lsnd}, which claims to have
measured a small mixing between the muon and electron anti-neutrinos.
The above assumption for the mass ranges of the neutrinos is also
quite natural in view of the mass hierarchy between generations
for the quarks and leptons. It is also compatible with the three
experimental neutrino anomalies (solar, atmospheric and LSND).
[Of course it is not the only possibility, but represents 
our best guess given the existing information].

For ordinary - sterile or ordinary - mirror neutrino 
two state mass mixing, the
weak-eigenstates ($\nu_{\alpha}, \nu_{s}$) will be 
linear combinations of two mass eigenstates ($\nu_a, \nu_b$):
\begin{equation}
\nu_{\alpha} = \cos\theta_0 \nu_a + \sin\theta_0 \nu_b,
\ \nu_{s} = -\sin\theta_0 \nu_a + \cos\theta_0 \nu_b.
\end{equation}
Note we will always define $\theta_0$ in such a way so
that $\cos 2\theta_0 \ge 0$.  We also adopt the convention that 
$\delta m^2 \equiv m_b^2 - m_a^2$. Hence 
with this convention $\delta m^2$ is positive 
(negative) provided that $m_b > m_a$ ($m_b < m_a$). 

Ordinary - sterile or ordinary - mirror neutrino oscillations can 
generate significant lepton number in the early Universe\cite{ftv, shi,
fv2}. The origin of this phenomenon can be traced to the fact that 
the effective potential induced from the coherent forward scattering 
of neutrinos with the background is generally unequal
to the effective potential for anti-neutrinos if the background
is CP asymmetric\cite{nr, msw}. This means that the matter mixing angles
for neutrinos are generally unequal to the matter mixing angles
for anti-neutrinos, and thus the oscillation rates for neutrino 
oscillations need not be the same as the oscillation rates for 
anti-neutrino oscillations.

We begin by briefly reviewing the case of ordinary -
sterile neutrino oscillations as developed in Ref.\cite{fv2}.
The evolution of lepton number 
$L_{\nu_{\alpha}} \equiv (n_{\nu_{\alpha}} - n_{\bar \nu_{\alpha}})/
n_{\gamma}$ (where the $n's$ are the number densities and 
$\alpha = e, \mu, \tau$) in the early Universe generated by
$\nu_{\alpha} - \nu_s$ oscillations can be approximately 
described  by the following equation\cite{fv2},
\begin{equation}
{dL_{\nu_{\alpha}} \over dt} =
{\pi^2 \over 4\zeta(3)T^3} \int {s^2 \Gamma^p_{\nu_{\alpha}} 
a^p (c - b^p)
(dn_{\nu_{\alpha}}^{+} - dn_{\nu_{s}}^{+})
\over [x^p + (c - b^p + a^p)^2][x^p + (c - b^p -a^p)^2] } + 
\Delta, 
\label{eeqq3}
\end{equation}
where $\Delta$ is a small correction term,
\begin{equation}
\Delta \simeq 
{-\pi^2 \over 8\zeta(3)T^3} \int {s^2 \Gamma^p_{\nu_{\alpha}} 
[x^p + (a^p)^2 + (b^P - c)^2] 
(dn_{\nu_{\alpha}}^{-} - dn_{\nu_{s}}^{-})
\over [x^p + (c - b^p + a^p)^2][x^p + (c - b^p -a^p)^2] }.  
\label{Delta}
\end{equation}
In these equations, $\zeta (3)$ is the Riemann zeta function 
of 3 ($\zeta (3) \simeq 1.202$),
$dn_{\nu_{\alpha}}^{\pm} \equiv dn_{\nu_{\alpha}} \pm 
dn_{\bar \nu_{\alpha}} $,
and similarly for $dn^{\pm}_{\nu_{s}}$.
In the region where the lepton number is much less than 1,
or equivalently, the chemical potential can be neglected,
\begin{equation}
dn_{\nu_{\alpha}} \simeq dn_{\bar \nu_{\alpha}} 
\simeq {1 \over 2\pi^2}{p^2 dp \over 1 + e^{p/T}}.
\end{equation}
In Eq.(\ref{eeqq3}, \ref{Delta}),  
$c \equiv \cos 2\theta_0,\ s \equiv \sin 2\theta_0$,
and the quantities, $b^p, a^p, x^p, \Gamma^p_{\nu_{\alpha}}$ 
are all functions of momentum of the form:
\begin{eqnarray}
& x^p = s^2 + {\Gamma^{2}_{\nu_{\alpha}} \over 4(\Delta_0^p)^2}
\left({p \over \langle p \rangle}\right)^2,
\ \Gamma^p_{\nu_{\alpha}} = \Gamma_{\nu_{\alpha}} 
{p \over \langle p \rangle}, \nonumber \\
& a^p = {-\sqrt{2}G_F n_{\gamma}L^{(\alpha)} \over \Delta_0^p},\  
b^p = {-\sqrt{2}G_F n_{\gamma} A_{\alpha}T^2\over \Delta_0^p M_W^2}
{p \over \langle p \rangle},
\label{ab}
\end{eqnarray}
where $n_{\gamma} = 2\zeta (3) T^3/\pi^2$, $\Delta_0^p = \delta m^2/2p$ 
and the thermally averaged collision frequencies 
$\Gamma_{\nu_{\alpha}}$ are 
\begin{equation}
\Gamma_{\nu_{\alpha}} \simeq y_{\alpha}G_F^2 T^5,
\label{Gammas}
\end{equation}
with $y_e \sim 4.0, y_{\mu, \tau} \simeq 2.9$\cite{B}.
In Eq.(\ref{ab}), $G_F$ is the Fermi constant, $M_W$ is 
the W-boson mass and $A_e \simeq 55.0, \ A_{\mu,\tau} 
\simeq 15.3$.  The function $L^{(\alpha)}$ is given by
\begin{equation}
L^{(\alpha)} = L_{\nu_{\alpha}} + L_{\nu_e} + L_{\nu_{\mu}}
+ L_{\nu_{\tau}} + \eta,
\label{Lsuper}
\end{equation}
where $\eta$ is a small asymmetry term which arises from the asymmetries 
of baryons and electrons and is expected to be about $10^{-10}$\cite{nr}.
With the definitions Eq.(\ref{ab}) observe that Eq.(\ref{eeqq3}) implies 
that significant lepton number can only be generated provided that 
$\delta m^2 < 0$ and in this case only for oscillations 
with $b^p < c$.

In Ref.\cite{fv2}, we showed that the distribution
of sterile states is governed approximately by
the equations
\begin{equation}
{dz \over dt} = {(1 - z) \over 4}{\Gamma_{\nu_{\alpha}}^p s^2 
\over x^p + (c - b^p + a^p)^2},\
{d\bar z \over dt} = {(1 - \bar z) \over 4}{\Gamma_{\nu_{\alpha}}^p s^2 
\over x^p + (c - b^p - a^p)^2},
\label{yq}
\end{equation}
where 
\begin{equation}
z \equiv {dn_{\nu_s}/dp \over dn_{\nu_{\alpha}}/dp},\
\bar z \equiv {dn_{\bar \nu_s}/dp \over dn_{\bar \nu_{\alpha}}/dp}.
\end{equation}
Thus, the equation governing the evolution
of $L_{\nu_{\alpha}}$ has the approximate form, 
\begin{equation}
{dL_{\nu_{\alpha}} \over dt} = 
{1 \over 4\zeta(3)T^3} \int^{\infty}_{0} {s^2 \Gamma^p_{\nu_{\alpha}} 
a^p (c - b^p)\over
[x^p + (c - b^p + a^p)^2][x^p + (c - b^p -a^p)^2]
} {(1 - z^+)p^2 dp \over (1 + e^{p/T})} + \Delta,
\label{eeqq4}
\end{equation}
where $\Delta$ is a small correction term,
\begin{equation}
\Delta \simeq
{1 \over 8\zeta(3)T^3} \int^{\infty}_{0} {s^2 \Gamma^p_{\nu_{\alpha}} 
[x^p + (a^p)^2 + (b^p - c)^2]\over
[x^p + (c - b^p + a^p)^2][x^p + (c - b^p -a^p)^2]
} {z^- p^2 dp \over (1 + e^{p/T})},
\end{equation}
with $z^{\pm} \equiv (z \pm \bar z)/2$ and
we have neglected a small term proportional to $L_{\nu_{\alpha}}$.

In the case of ordinary - mirror neutrino oscillations,
there are two important modifications. First, the effective
potential of the mirror neutrinos is generally non-negligible.
Second, the mirror weak interactions can bring the mirror
neutrino into thermal equilibrium with other light mirror
particles. We now discuss these points in more detail.

In the case of $\nu_{\alpha} - \nu'_{\beta}$ oscillations,
the dynamics depend on the difference of the effective
potentials,
\begin{equation}
V = V_{\alpha} - V_{\beta}',
\end{equation}
where $V_{\alpha}$ ($V'_{\beta}$) is the effective potential
experienced by a pure weak (mirror) eigenstate.
These effective potentials can be expressed in terms of
the parameters $a^p, b^p$ ($a'^p, b'^p$) as follows,
\begin{equation}
V_{\alpha} = (-a^p + b^p)\Delta_0^p,\
V'_{\beta} = (-a'^p + b'^p)\Delta_0^p.
\end{equation}
If the number of mirror neutrinos is much less than the
number of ordinary neutrinos then $b'^p \simeq 0$.
[Note that the $b$-part of the effective potential is
proportional to the number densities of the background particles.
This dependence is not given explicitly in Eq.(\ref{ab}) since for this
equation the number densities were set equal to their
equilibrium values].  The parameter $a'^p$ has the form
\begin{equation}
a'^p \equiv {-\sqrt{2}G_F n_{\gamma}L'^{(\beta)} \over \Delta_0^p},  
\end{equation}
where
$L'^{(\beta)}$ is given by 
\begin{equation}
L'^{(\beta)}  = L_{\nu'_{\beta}} + L_{\nu'_e} + L_{\nu'_{\mu}}
+ L_{\nu'_{\tau}} + \eta',
\end{equation}
and $L_{\nu'_{\beta}}$ are the mirror lepton numbers, which
are defined by $L_{\nu'_{\beta}} \equiv  (n_{\nu_{\beta}'} 
- n_{\bar \nu'_{\beta}})/n_{\gamma}$ (note that $n_{\gamma}$ 
is the number density of {\it ordinary} photons) and 
$\eta'$ is a function of the mirror baryon/electron number asymmetries
[which is defined analogously to $\eta$].  We
will assume that $\eta'$ is small and can be approximately neglected.
Thus, assuming that the number density of mirror
particles is much less than the number density of ordinary particles,
the modification of the effective potential due to the
mirror interactions of $\nu_{\beta}'$ can
be approximately taken into account by simply replacing
$L^{(\alpha)}$ in the definition of $a^p$ by 
$\stackrel{\sim}{L}^{\alpha \beta'} \equiv L^{(\alpha)} - 
L'^{(\beta)}$.  Since ordinary + mirror lepton number is conserved 
(and we will assume that it is zero), it follows that
\begin{equation}
L_{\nu_e} + L_{\nu_{\mu}} + L_{\nu_{\tau}} + 
L_{\nu'_e} + L_{\nu'_{\mu}} + L_{\nu'_{\tau}} = 0.
\label{145}
\end{equation}
For example, if we consider
$\nu_{\alpha} - \nu'_{\beta}$ oscillations
in isolation then $L_{\nu'_{\beta}} = -L_{\nu_{\alpha}}$ and
$\stackrel{\sim}{L}^{\alpha \beta'} \simeq 2L^{(\alpha)}$. 

Another important effect of the mirror interactions is that
the momentum distribution of the mirror neutrinos will
be approximately Fermi-Dirac distributions, and 
the other mirror particles will be excited until 
\begin{equation}
T_{\nu_e'} = T_{\nu'_{\mu}} = T_{\nu'_{\tau}} = T_{e'} = T_{\gamma'} 
\equiv T',
\end{equation}
(we consider the $T' \stackrel{<}{\sim} 100 \ {\rm MeV}$
region only where we can approximately neglect the excitation
of mirror muons and other heavier mirror particles).
Of course the light mirror particles will only be excited
provided that there 
are sufficient mirror neutrinos around so that the interaction
rates will be faster than the expansion rate, that is, 
\begin{equation}
hG_F^2 T'^5 \stackrel{>}{\sim} 5.5T^2/M_P,
\label{y2}
\end{equation}
where $M_P$ is the Planck mass and $h$ is a numerical parameter 
which depends on the particular
interaction (see Table 1 of Ref.\cite{ekm} for a list of the
reaction rates).  Typically, $h \sim 1/8$.
Solving Eq.(\ref{y2}), we obtain the condition
\begin{equation}
T'\stackrel{>}{\sim} 2\left({T \over {\rm MeV}}\right)^{2 \over 5} 
\ {\rm MeV}.
\label{y3}
\end{equation}
Assuming that the above condition is approximately satisfied, the 
system of ordinary and mirror particles form two weakly coupled 
thermodynamic systems; the system comprising the ordinary
particles at a temperature $T$, and the system comprising the
mirror particles which has a distinct temperature $T'$.

Let us determine the equation governing the evolution
of $T'$. Initially, we will assume that $T' = 0$.
Ordinary - mirror neutrino oscillations can then
generate a mirror neutrino $\nu'_{\beta}$ say.
The rate at which mirror neutrinos are created/destroyed by
$\nu_{\alpha} - \nu'_{\beta}$ oscillations is governed by the
rate equation,
\begin{equation}
{d \over dt}\left[ {n_{\nu'_{\beta}} + n_{\bar \nu'_{\beta}} \over
n_{\nu_{\alpha}} + n_{\bar \nu_{\alpha}}} \right] = 
{1 \over n_{\nu_{\alpha}} + n_{\bar \nu_{\alpha}}}\int
\Gamma (\nu_{\alpha} \to \nu'_{\beta})(dn_{\nu_{\alpha}} - 
dn_{\nu'_{\beta}}) + \Gamma (\bar \nu_{\alpha} \to \bar \nu'_{\beta})
(dn_{\bar \nu_{\alpha}} - dn_{\bar \nu'_{\beta}}). 
\label{mon}
\end{equation}
The reaction rates are given by\cite{fv2,qqq},
\begin{equation}
 \Gamma (\nu_{\alpha} \to \nu'_{\beta}) = {1 \over 2}\Gamma^p_{\nu_{\alpha}}
\sin^2 2\theta_m \langle \sin^2 {\tau \over 2L_{osc}^{m}}\rangle
 = {1 \over 4}\left[ { \Gamma^p_{\nu_{\alpha}} 
s^2 \over x^p + (b^p - a^p - c)^2}\right].
\label{rates}
\end{equation}
The rate for for the process $\Gamma (\bar \nu_{\alpha} \to
\bar \nu'_{\beta})$ can be obtained by replacing $a^p \to -a^p$ in the above
equation.  
If Eq.(\ref{y3}) is satisfied, then the momentum distribution
of sterile neutrinos has the form
\begin{equation}
dn_{\nu'_{\beta}} \simeq dn_{\bar \nu'_{\beta}} \simeq
{1 \over 2\pi^2}{p^2 dp \over 1 + e^{p/T'}},
\label{x2}
\end{equation}
where we have neglected the mirror neutrino chemical potential
which is small provided that the lepton number is sufficiently small.
Thus using Eqs.(\ref{rates}) and Eq.(\ref{x2}),
it is straightforward to show that 
Eq.(\ref{mon}) can be expressed as follows,
\begin{equation}
{d \over dt}\left[{n_{\nu'_{\beta}} + n_{\bar \nu'_{\beta}} \over
n_{\nu_{\alpha}} + n_{\bar \nu_{\alpha}}} \right] \simeq 
{1 \over 6\zeta (3)T^3}\int^{\infty}_0 {s^2 \Gamma_{\nu_{\alpha}}^p 
[(b^p - c)^2 + (a^p)^2 + x^p]p^2 F(p,T,T') dp \over
[x^p + (c - b^p + a^p)^2][x^p + (c - b^p -a^p)^2]},
\label{yyy}
\end{equation}
where 
\begin{equation}
F(p,T,T') \equiv {1 \over 1 + e^{p/T}} - {1 \over 1 + e^{p/T'}}.
\end{equation}
During the process whereby the mirror interactions excite the
other mirror particles and thermalize the momentum distributions,
the number density of sterile neutrinos is generally not conserved. 
Thus, we cannot directly use Eq.(\ref{yyy}) to determine the evolution 
of $T'$.  However, the mirror interactions must conserve energy.
If the energy density of the mirror particles is much less than the
energy density of the ordinary particles then the process
whereby the mirror sector comes into equilibrium with itself should
not significantly affect the expansion rate of the Universe.
For this reason, and because of the conservation 
of energy, it follows that the energy density of the mirror particles 
normalized to the energy density of the ordinary particles
should to a good approximation not change due to the 
expansion of the Universe.  We will denote this quantity by
$\gamma \equiv \rho'/\rho$. In the region where
$1 \ {\rm MeV} \stackrel{<}{\sim} T \stackrel{<}{\sim} 100 \ {\rm MeV}$,
and $1 \ {\rm MeV} \stackrel{<}{\sim} T' \stackrel{<}{\sim} 
100 \ {\rm MeV}$, then $\gamma \simeq T'^4/T^4$
(assuming that we are in a region where Eq.(\ref{y3}) is valid).
Using Eq.(\ref{yyy}), it is straightforward to show that the 
evolution of $\gamma$, which is entirely due to ordinary - mirror
transitions, satisfies the following equation (where we are
assuming that $\gamma \ll 1$),
\begin{equation}
{d\gamma \over dt} \simeq {1 \over 19\zeta (3)NT^4}\int^{\infty}_0
{s^2 \Gamma_{\nu_{\alpha}}^p 
[(b^p - c)^2 + (a^p)^2 + x^p] p^3 F(p,T,T')dp\over
[x^p + (c - b^p + a^p)^2][x^p + (c - b^p -a^p)^2]},
\label{y5}
\end{equation}
where $N = \rho'/\rho'_{\nu'}$.
Using $\rho'= 3\rho'_{\nu'} + \rho'_{e'} + \rho'_{\gamma'} 
= 6{1 \over 7} \rho'_{\nu'}$, it follows that $N \simeq 6.14$.

In the region where Eq.(\ref{y3}) holds, 
Eq.(\ref{eeqq3}) and Eq.(\ref{x2}) imply that
the evolution of lepton number obeys the following equation,
\begin{equation}
{dL_{\nu_{\alpha}} \over dt} =
{1 \over 4\zeta(3)T^3} \int^{\infty}_{0} {s^2 \Gamma^p_{\nu_{\alpha}} 
a^p (c - b^p)p^2 F(p,T,T')dp\over
[x^p + (c - b^p + a^p)^2][x^p + (c - b^p -a^p)^2] } + \Delta, 
\label{y6}
\end{equation}
where\cite{hhh}
\begin{equation}
\Delta \simeq
{-3L_{\nu_{\alpha}} \over 2\pi^2} \int^{\infty}_{0} 
{s^2 \Gamma^p_{\nu_{\alpha}} [x^p + (a^p)^2 + (b^p - c)^2]
G(p,T,T')dp\over
[x^p + (c - b^p + a^p)^2][x^p + (c - b^p -a^p)^2] },
\end{equation}
and 
\begin{equation}
G(p,T,T') \equiv {1 \over T^3}{p^2 e^{p/T} \over (1 + e^{p/T})^2}
- {1 \over T'^3}{p^2 e^{p/T'} \over (1 + e^{p/T'})^2}.
\end{equation}
Equations (\ref{y5}) and (\ref{y6}) are coupled differential equations
that can be solved for $L_{\nu_{\alpha}}, T'$.

In the region where 
\begin{equation}
T' \ll 2\left({T \over {\rm MeV}}\right)^{2 \over 5} 
\ {\rm MeV},
\label{i3}
\end{equation}
the number of mirror particles are insufficient to
enable the mirror sector to come into equilibrium with
itself. 
In the case of Eq.(\ref{i3}), the momentum
distribution and number density of mirror neutrinos are
not significantly affected by the mirror interactions.
This
means that the momentum distribution of mirror neutrinos
should be the same as with the case of sterile neutrinos.
Thus the momentum distribution and number
density of mirror neutrinos should be governed 
approximately by Eq.(\ref{yq}) with
the evolution of
lepton number governed approximately by 
Eq.(\ref{eeqq4}). 
Of course even in the case of Eq.(\ref{i3}), the effects of 
the mirror interactions on the
effective potential must still be taken into account,
that is, we must replace $L^{(\alpha)}$ by 
$\stackrel{\sim}{L}^{\alpha \beta'}$.

There are several ways in which the creation of lepton number(s)
can prevent sterile/mirror neutrinos from coming into
equilibrium. One way is that one set of oscillations $\nu_{\alpha}-
\nu_{s}$ creates $L_{\nu_{\alpha}}$. The lepton number
$L_{\nu_{\alpha}}$ can then suppress other, independent
oscillations such as $\nu_{\beta} - \nu_{s}$ oscillations (with
$\beta \neq \alpha$) for example.  A more direct, but less dramatic 
way in which the creation of lepton number can help
prevent the sterile/mirror neutrinos from coming into
equilibrium, is that the lepton number generated from
say $\nu_{\alpha}-\nu_{s}$ oscillations  itself suppresses the 
$\nu_{\alpha}-\nu_{s}$ oscillations\cite{kc}. 
We will examine the latter effect here (the former
effect will be studied in a moment). Previous work\cite{B}
obtained the BBN bound on ordinary - sterile neutrino oscillations
with large $|\delta m^2| \stackrel{>}{\sim} 
10^{-4} \ eV^2$ (with $\delta m^2 < 0$) and small 
$\sin^2 2\theta_0 \stackrel{<}{\sim} 10^{-2}$. This bound can be
approximately parametrized as follows
\begin{equation}
\sin^2 2\theta_0 \stackrel{<}{\sim} 3\times 10^{-5} \left(
{ eV^2 \over |\delta m^2|}\right)^{1 \over 2}.
\label{blob}
\end{equation} 
This bound arises by assuming that the 
$\nu_{\alpha} - \nu_{s}$ oscillations do not bring the sterile
$\nu_s$ state into equilibrium. Note that this bound neglected
the creation of lepton number and it also did not include the effects
of the distribution of neutrino momenta. However, in the realistic 
case, the creation of $L_{\nu_{\alpha}}$ (after it occurs), will 
suppress the $\nu_{\alpha} - \nu_{s}$ oscillations and the actual 
bound might be expected to be somewhat less stringent than Eq.(\ref{blob}).
Nevertheless, for the case of truly sterile neutrinos we found\cite{fv2} that
Eq.(\ref{blob}) turned out to be a reasonable approximation in 
the realistic case where the momentum distribution and the
creation of lepton number was taken into account.
This is largely due to the fact that the creation of sterile neutrinos 
suppresses the lepton number creating oscillations
and consequently delays the point where significant
lepton number can be created\cite{fv2}.
In the case of mirror neutrinos, 
there will be much less of this type of effect
because the number density of the mirror neutrinos
is kept low due to the excitation of the other mirror particles.
Also, the thermalization of the neutrino momentum distributions
means that not all of the mirror neutrinos will have low momentum.
Thus, we would expect that the bound Eq.(\ref{blob}) should be 
weakened somewhat.

For the case of ordinary - mirror neutrino oscillations,
the BBN bound for large $|\delta m^2| \stackrel{>}{\sim} 
10^{-4} \ eV^2$ (with $\delta m^2 < 0$) and small 
$\sin^2 2\theta_0 \stackrel{<}{\sim} 10^{-2}$ can be
obtained by solving the coupled differential equations,
Eq.(\ref{y5}) and Eq.(\ref{y6}).
Doing this numerical exercise, we obtain the following 
bounds assuming that $\delta N_{eff} = \rho'/\rho_{\nu_{\alpha}}
 < 0.6 \ (1.5)$\cite{bbn} (where $N_{eff} $ is the effective
number of neutrinos present during nucleosynthesis),
\begin{eqnarray}
&\sin^2 2\theta_0 \stackrel{<}{\sim} 3(6)\times 10^{-4} \left(
{ eV^2 \over |\delta m^2|}\right)^{1 \over 2} \ {\rm for } \
\nu_{\mu, \tau} - \nu'_{\beta} \ {\rm oscillations}, \nonumber \\
&\sin^2 2\theta_0 \stackrel{<}{\sim} 1(2)\times 10^{-4} \left(
{ eV^2 \over |\delta m^2|}\right)^{1 \over 2} \ {\rm for } \
\nu_{e} - \nu'_{\beta} \ {\rm oscillations}.
\label{blob2}
\end{eqnarray} 
Comparing these bounds with Eq.(\ref{blob}), we see that the BBN bound 
on $\sin^2 2\theta_0$ for ordinary - mirror neutrino oscillations 
is about an order of magnitude weaker than the corresponding 
bound for ordinary - sterile neutrino oscillations.

We now identify the region of parameter space
for which the  maximal ordinary - mirror neutrino oscillations
can solve the solar and atmospheric neutrino anomalies
without leading to a significant modification of BBN.

We first study the maximal ordinary - mirror neutrino
oscillation solution to the atmospheric neutrino problem.
We will assume that the various oscillations can be approximately
broken up into the pairwise oscillations $\nu_{\alpha} - \nu'_{\beta}$.
We will denote the various oscillation parameters in a self-evident
notation,
\begin{equation}
b^p_{\alpha \beta'}, a^p_{\alpha \beta'} \ {\rm for \ } \nu_{\alpha} 
- \nu'_{\beta} \ {\rm oscillations,\ }
\end{equation}
where $\alpha, \beta = e, \mu, \tau$.
We will denote the mixing parameters, $\delta m^2, \ \sin^2 2\theta_0$ 
appropriate for $\nu_{\alpha} - \nu'_{\beta}$ oscillations by
$\delta m^2_{\alpha \beta'}, \ \sin^2 2\theta_0^{\alpha \beta'}$.
Note that lepton number cannot be created by $\nu_{\alpha} 
- \nu'_{\beta}$ oscillations until $\langle b^p_{\alpha \beta'} 
\rangle \stackrel{<}{\sim} \cos2\theta_0^{\alpha \beta'}$.  Recall 
that the $b^p$ 
parameter is inversely proportional to $\delta m^2$ [see Eq.(\ref{ab})].
Thus, the earliest point during the evolution of the Universe
where lepton number can be created due to ordinary - mirror
neutrino oscillations occurs for oscillations which have the largest
$|\delta m^2|$. Note that these oscillations 
should satisfy the BBN bound Eq.(\ref{blob2}) if we
demand that the sterile neutrino should not significantly modify
BBN. Note that the $\nu_{\mu} - \nu'_{\mu}$ oscillations have quite small
$|\delta m^2_{\mu \mu'}| \sim 10^{-2}\ eV^2$, and
$\cos 2\theta_0^{\mu \mu'} \sim 0$ (assuming maximal or
near maximal mixing), and thus these oscillations 
themselves cannot produce significant lepton number.
However, the $|\delta m^2|$ for $\nu_{\tau} - \nu'_{\mu}$
(or $\nu_{\tau} - \nu'_e$) oscillations
can be much larger. Also note that $\delta m^2 < 0$ if 
$m_{\nu_{\tau}} > m_{\nu'_{\mu}}$ (or $m_{\nu_{\tau}} > m_{\nu'_e}$).

We will first consider the system comprising  
$\nu_{\tau}, \nu_{\mu}$ and $\nu'_{\mu}$ (and their anti-particles).
The effects of the other neutrinos will be discussed in a moment.
We will assume for definiteness that $m_{\nu_{\tau}} > m_{\nu'_{\mu}}$
so that $|\delta m^2_{\tau \mu'}| > |\delta m^2_{\mu \mu'}|$ 
and the $\nu_{\tau} - \nu'_{\mu}$ oscillations create $L_{\nu_{\tau}}$
first (with $L_{\nu_{\mu}}, L_{\nu_e}$ assumed to be initially
negligible). 

Recall that in the case of $\nu_{\alpha} - \nu'_{\beta}$ oscillations,
the effect of the mirror interactions on the effective
potential can be taken into account by replacing 
$L^{(\alpha)}$ in the definition of $a^p$ by 
$\stackrel{\sim}{L}^{\alpha  \beta'} = L^{(\alpha)} - L'^{(\beta)}$.
For the $\nu_{\tau}, \nu_{\mu}, \nu'_{\mu}$ system,
\begin{eqnarray}
& \stackrel{\sim}{L}^{\mu \mu'} \simeq
2L_{\nu_{\mu}} + L_{\nu_{\tau}} - 2L_{\nu'_{\mu}} \simeq
4L_{\nu_{\mu}} + 3L_{\nu_{\tau}}, \nonumber \\
& \stackrel{\sim}{L}^{\tau \mu'} \simeq
2L_{\nu_{\tau}} + L_{\nu_{\mu}} - 2L_{\nu'_{\mu}} \simeq
4L_{\nu_{\tau}} + 3L_{\nu_{\mu}},
\end{eqnarray}
where we have used Eq.(\ref{145}) with $L_{\nu_e} \simeq L_{\nu_e'} 
\simeq L_{\nu'_{\tau}} \simeq 0$.
Thus, the creation of $L_{\nu_{\tau}}$ by $\nu_{\tau} - \nu'_{\mu}$
oscillations also implies the creation of the quantity
$\stackrel{\sim}{L}^{\mu \mu'}$. 
Because $a^p_{\mu \mu'}$ is directly proportional to 
$\stackrel{\sim}{L}^{\mu \mu'}$ it follows that the creation
of a large $\stackrel{\sim}{L}^{\mu \mu'}$ will make $a^p_{\mu \mu'}$
large (i.e. significantly greater than 1), which thereby
suppresses the oscillations (note that $\sin^2 2\theta_m^{\mu \mu'}
\sim \sin^2 2\theta_0^{\mu \mu'}/a^2_{\mu \mu'}$ if 
$a_{\mu \mu'} \gg 1$).
However, note that $\nu_{\mu} - \nu'_{\mu}$ oscillations
can potentially destroy $\stackrel{\sim}{L}^{\mu \mu'}$
(because for these oscillations $\stackrel{\sim}{L}^{\mu \mu'} 
\simeq 0$ is an approximately stable fixed point for the 
temperature range of interest).  
Thus, we must obtain the region of parameter space where 
$\stackrel{\sim}{L}^{\mu \mu'}$ created by $\nu_{\tau}
- \nu'_{\mu}$ oscillations does not get subsequently destroyed
by $\nu_{\mu} - \nu'_{\mu}$ oscillations.  We will determine this 
region of parameter space by numerically solving the coupled
differential equations governing the evolution of
$L_{\nu_{\tau}}, L_{\nu_{\mu}}$ and $T'$.
[An approximate analytic computation could be done
following similar reasoning to the case of
sterile neutrinos\cite{fv2}. Note however that the approximations used
in this analytic derivation are not valid if
the point where the destruction of $\stackrel{\sim}{L}^{\mu \mu'}$ 
due to $\nu_{\mu} - \nu'_{\mu}$ oscillations
reaches a maximum can occur during the phase where the
creation of $\stackrel{\sim}{L}^{\mu \mu'}$ due to
$\nu_{\tau} - \nu'_{\mu}$ oscillations is still growing
exponentially].

The rate of change of $L_{\nu_{\mu}}$ and $L_{\nu_{\tau}}$ due
to the $\nu_{\tau} - \nu'_{\mu}$, $\nu_{\mu} - \nu'_{\mu}$
oscillations can be obtained from Eq.(\ref{y6}).
This leads to the following coupled differential equations,
\begin{eqnarray}
& {dL_{\nu_{\mu}} \over dt} \simeq 
{1 \over 4\zeta(3)T^3} \int^{\infty}_{0} {\sin^2 2\theta_0^{\mu \mu'}
\Gamma^p_{\nu_{\mu}} a^p_{\mu \mu'}  (\cos2\theta_0^{\mu \mu'} -
b^p_{\mu \mu'})p^2 F(p,T,T') dp\over
[x^p_{\mu \mu'} + (\cos2\theta_0^{\mu \mu'}- b^p_{\mu \mu'} + 
a^p_{\mu \mu'})^2][x^p_{\mu \mu'} + (\cos2\theta_0^{\mu \mu'} - 
b^p_{\mu \mu'}  - a^p_{\mu \mu'})^2]
} \nonumber \\
& - {3L_{\nu_{\mu}} \over 2\pi^2} \int^{\infty}_{0} 
{\sin^2 2\theta_0^{\mu \mu'} \Gamma^p_{\nu_{\mu}} 
[x^p_{\mu \mu'} + (a^p_{\mu \mu'})^2 + (b^p_{\mu \mu'} - 
\cos2\theta_0^{\mu \mu'})^2] G(p,T,T')dp\over
[x^p_{\mu \mu'} + (\cos2\theta_0^{\mu \mu'}- b^p_{\mu \mu'} + 
a^p_{\mu \mu'})^2][x^p_{\mu \mu'} + (\cos2\theta_0^{\mu \mu'} - 
b^p_{\mu \mu'}  - a^p_{\mu \mu'})^2]
}, \nonumber \\
& {dL_{\nu_{\tau}} \over dt} \simeq 
{1 \over 4\zeta(3)T^3} \int^{\infty}_{0} {\sin^2 2\theta_0^{\tau \mu'}
\Gamma^p_{\nu_{\tau}} 
a^p_{\tau \mu'}  (\cos2\theta_0^{\tau \mu'} -b^p_{\tau \mu'})p^2 
F(p,T,T')dp\over
[x^p_{\tau \mu'} + (\cos2\theta_0^{\tau \mu'}- b^p_{\tau \mu'} + 
a^p_{\tau \mu'})^2][x^p_{\tau \mu'} + (\cos2\theta_0^{\tau \mu'} - 
b^p_{\tau \mu'}  - a^p_{\tau \mu'})^2]}
\nonumber \\
& - {3L_{\nu_{\tau}} \over 2\pi^2} \int^{\infty}_{0} 
{\sin^2 2\theta_0^{\tau \mu'} \Gamma^p_{\nu_{\tau}} 
[x^p_{\tau \mu'} + (a^p_{\tau \mu'})^2 + (b^p_{\tau \mu'} - 
\cos2\theta_0^{\tau \mu'})^2] G(p,T,T')dp\over
[x^p_{\tau \mu'} + (\cos2\theta_0^{\tau \mu'}- b^p_{\tau \mu'} + 
a^p_{\tau \mu'})^2][x^p_{\tau \mu'} + (\cos2\theta_0^{\tau \mu'} - 
b^p_{\tau \mu'}  - a^p_{\tau \mu'})^2]
}, 
\label{longeq2}
\end{eqnarray}
where the equation governing the evolution of $T'$ can be obtained
from Eq.(\ref{y5}), and it is
\begin{eqnarray}
&{d\gamma \over dt} \simeq {1 \over 19\zeta (3) NT^4}\int^{\infty}_0
{\sin^2 2\theta_0^{\mu \mu'} \Gamma_{\nu_{\mu}}^p 
[(b^p_{\mu \mu'} - \cos2\theta_0^{\mu \mu'})^2 + (a^p_{\mu \mu'})^2 + 
x^p_{\mu \mu'}] p^3 F(p,T,T')dp\over
[x^p_{\mu \mu'} + (\cos2\theta_0^{\mu \mu'} - b^p_{\mu \mu'} + 
a^p_{\mu \mu'})^2][x^p_{\mu \mu'} + (\cos2\theta_0^{\mu \mu'} - 
b^p_{\mu \mu'} -a^p_{\mu \mu'})^2]}
\nonumber \\
& + {1 \over 19\zeta (3) NT^4}\int^{\infty}_0
{\sin^2 2\theta_0^{\tau \mu'} \Gamma_{\nu_{\tau}}^p 
[(b^p_{\tau \mu'} - \cos2\theta_0^{\tau \mu'})^2 + (a^p_{\tau \mu'})^2 + 
x^p_{\tau \mu'}]p^3 F(p,T,T')dp \over
[x^p_{\tau \mu'} + (\cos2\theta_0^{\tau \mu'} - b^p_{\tau \mu'} + 
a^p_{\tau \mu'})^2][x^p_{\tau \mu'} + (\cos2\theta_0^{\tau \mu'} - 
b^p_{\tau \mu'} -a^p_{\tau \mu'})^2]}.
\label{y7}
\end{eqnarray}
Note that Eq.(\ref{y7}) and Eq.(\ref{longeq2}) are coupled
differential equations which must be solved simultaneously.
In our numerical integration
of Eqs.(\ref{longeq2}, \ref{y7}), we will assume for 
definiteness that $\sin^2 2\theta_0^{\mu \mu'}$  and
that $\delta m^2_{\mu \mu'}$ are given by the best fit for the 
atmospheric neutrino data, i.e. $\sin^2 2\theta_0^{\mu \mu'} \simeq 1$
and $|\delta m^2_{\mu \mu'}| \simeq 10^{-2} \ eV^2$.  
The results of this exercise are shown 
in Figure 1 \cite{remark, remark2}.  In the region above the solid line, the 
$\stackrel{\sim}{L}^{\mu \mu'}$ 
created by $\nu_{\tau} - \nu'_{\mu}$ oscillations does not
get subsequently destroyed by $\nu_{\mu} - \nu'_{\mu}$ oscillations.
The numerical work also demonstrates that
the lepton number is generated early enough
and is large enough to suppress the $\nu_{\mu} - \nu'_{\mu}$
oscillations sufficiently so that these oscillations do not
create a significant number of $\nu'_{\mu}$ states.
The requirement that $\nu_{\tau} - \nu'_{\mu}$ oscillations
do not produce too many sterile states 
implies an upper limit on $\sin^2 2\theta_0^{\tau \mu'}$ [see
Eq.(\ref{blob2})]. This upper limit has been shown in the Figure
assuming for definiteness that $\rho'/\rho_{\nu_{\alpha}} < 0.6$ 
(dashed-dotted line). Also shown in Figure 1 (dashed line) is 
the cosmological energy density  bound $|\delta m^2_{\tau \mu'}| 
\stackrel{<}{\sim} 1600 \ eV^2$\cite{fn39}.

Comparing the allowed region of parameter space shown in Figure 1
with the analogous case for sterile neutrinos\cite{fv2}, it is
apparent that the corresponding allowed region for mirror neutrinos is
somewhat larger than the allowed region for sterile neutrinos.
The increase of parameter space (which is about
an order of magnitude larger for $\sin^2 2\theta_0^{\tau \mu'}$) 
in the case
of mirror neutrinos is primarily due to the result 
that the bound, Eq.(\ref{blob2}),
is considerably less stringent than the bound, Eq.(\ref{blob}).

While we have focussed on the $\stackrel{\sim}{L}^{\mu \mu'}$ 
generated by $\nu_{\tau} - \nu'_{\mu}$ oscillations,
similar results will also hold for $\stackrel{\sim}{L}^{\mu \mu'}$
generated by $\nu_{\tau} - \nu'_{e}$ oscillations.
Note that in this case,
\begin{eqnarray}
& \stackrel{\sim}{L}^{\mu \mu'} \simeq
2L_{\nu_{\mu}} + L_{\nu_{\tau}} - 2L_{\nu'_{\mu}} - L_{\nu'_{e}} \simeq
4L_{\nu_{\mu}} + 2L_{\nu_{\tau}}, \nonumber \\ 
& \stackrel{\sim}{L}^{\tau e'} \simeq
2L_{\nu_{\tau}} + L_{\nu_{\mu}} - 2L_{\nu'_{e}} - L_{\nu'_{\mu}}\simeq
4L_{\nu_{\tau}} + 2L_{\nu_{\mu}}.
\end{eqnarray}
In this case, solving the appropriate coupled differential
equations, we find that the $\stackrel{\sim}{L}^{\mu \mu'}$ generated
by $\nu_{\tau} - \nu'_{e}$ oscillations does not get destroyed
by $\nu_{\mu} - \nu'_{\mu}$ oscillations for the range of
parameter space shown in Figure 2.
Of course we only require that the oscillation parameters
are in the allowed region shown in Figure 1 {\it or} Figure 2.

Note that our results should only be considered as approximate
because we have only included the $\nu_{\tau} - \nu'_{\beta}$ 
and $\nu_{\mu} - \nu'_{\mu}$ oscillations (where $\beta = \mu$ for
Figure 1 and $\beta = e$ for Figure 2).
In the most general case, the system is considerably more 
complicated. In general one should also
include the effects of the oscillations with
$\delta m^2 > 0$ as well as ordinary - ordinary and
mirror - mirror neutrino oscillations.
The oscillations with $\delta m^2 > 0$,
include oscillations such as $\nu_{\mu} - \nu'_{\tau}$  
[given the assumption Eq.(\ref{hierarchy})]. 
Focusing on this example,
these oscillations generate $L_{\nu_{\mu}}, L_{\nu'_{\tau}}$ 
such that $L^{(\mu)} - L'^{(\tau)} \to 0$. 
They cannot prevent $\nu_{\tau} - \nu_{\mu}'$ or $\nu_{\tau} -
\nu'_{e}$ oscillations from generating $L_{\nu_{\tau}}$ and hence
$\stackrel{\sim}{L}^{\mu \mu'}$. For this reason these oscillations
cannot change anything qualitatively and for simplicity we have
neglected them.  The effect of ordinary - ordinary neutrino
oscillations, such as $\nu_{\tau} - \nu_{\mu}$ oscillations is to
generate $L_{\nu_{\mu}}, L_{\nu_{\tau}}$ such that $L_{\nu_{\mu}} 
- L_{\nu_{\tau}} \to 0$.  These oscillations also cannot prevent 
$\stackrel{\sim} {L}^{\mu \mu'}$ being generated by 
$\nu_{\tau} - \nu_{\mu}'$ (or $\nu_{\tau} - \nu'_{e}$) oscillations. 
Also the rate of change lepton number due to these 
oscillations is typically suppressed compared to the rate of 
change of lepton number due to ordinary - sterile (or mirror) 
neutrino oscillations\cite{shi, fv2}.
Finally, the effect of mirror - mirror neutrino oscillations
may also be important, although the precise effect
of these oscillations is less clear.

We turn to a brief discussion of the maximal oscillation
solution to the solar neutrino problem.
Recall that maximal $\nu_e - \nu'_e$ oscillations
can lead to a simple predictive solution to the
solar neutrino problem for the large range of
parameters, Eq.(\ref{solar}).
Note that much of this parameter space is naively in conflict
with BBN [see Eq.(\ref{B})].
However, $\nu_{\tau} - \nu'_e$ or $\nu_{\tau} - \nu_{\mu}'$ 
oscillations will generate $\stackrel{\sim}{L}^{ee'}$ which
can suppress the $\nu_e - \nu'_e$ oscillations
provided that $\nu_e - \nu'_e$ oscillations do not
destroy the $\stackrel{\sim}{L}^{ee'}$ asymmetry.
The situation is completely analogous to the case involving
the $\nu_{\mu} - \nu'_{\mu}$ oscillations that we
have been studying above.
For such large $|\delta m^2_{\tau \mu'}|$ or 
$|\delta m^2_{\tau e'}|$ identified in the Figures,
it turns out that essentially the entire parameter
space Eq.(\ref{solar}) does not lead to
any significant modification of BBN (which 
is largely due to the fact that $|\delta m^2_{ee'}| 
\stackrel{<}{\sim} |\delta m^2_{\mu \mu'}|$).

In summary, the solutions of the atmospheric and solar neutrino
problems suggested by the exact parity model are {\it not}
in conflict with BBN for a significant range of parameters.
Consistency with BBN does require relatively large
values of the parameters,
\begin{equation}
|\delta m^2_{\tau \mu'}|\ ({\rm or}\  |\delta m^2_{\tau e'}|)
\stackrel{>}{\sim} 25 \ eV^2 \ (50 \ eV^2).
\end{equation}
This suggests that $m_{\nu_{\tau}} \stackrel{>}{\sim} 5
\ eV$. Such values of the tau
mass can also be motivated independently by the evidence for 
dark matter.  Thus, demanding that the exact parity model solution
of the atmospheric neutrino anomaly be consistent with BBN
implies that the tau neutrino mass should be in a cosmologically
interesting range.
Also note that this tau neutrino mass range will be
probed by the NOMAD and CHORUS experiments\cite{nomad}.

\vskip 1cm
\noindent
{\large \bf Figure Captions}
\vskip 0.5cm
\noindent
Figure 1. Region of parameter space in the $\sin^2 2\theta_0^{\tau \mu'},
\ - \delta m^2_{\tau \mu'}$ plane where the $\stackrel{\sim}{L}^{\mu \mu'}$
created by $\nu_{\tau} - \nu'_{\mu}$ oscillations does not
get destroyed by $\nu_{\mu} - \nu'_{\mu}$ oscillations.
This region which in the Figure is denoted by the ``Allowed Region'',
includes all of the parameter space above the solid line. We have assumed
that $\sin^2 2\theta_0^{\mu \mu'} \simeq 1$ and $|\delta m^2_{\mu \mu'}|
= 10^{-2}\ eV^2$ (which is the best fit to the atmospheric neutrino
data). Also shown (dashed line) is the cosmology bound
$|\delta m^2_{\tau \mu'}| \stackrel{<}{\sim} 1600 \ eV^2$.
The dashed-dotted line is the nucleosynthesis bound
Eq.(\ref{blob2}) assuming that $ N_{eff} = \rho'/\rho_{\nu_{\alpha}}
\stackrel{<}{\sim} 0.6$.
\vskip 0.5cm
\noindent
Figure 2. Region of parameter space in the $\sin^2 2\theta_0^{\tau e'},
\ - \delta m^2_{\tau e'}$ plane where the $\stackrel{\sim}{L}^{\mu \mu'}$
created by $\nu_{\tau} - \nu'_{e}$ oscillations does not
get destroyed by $\nu_{\mu} - \nu'_{\mu}$ oscillations.  As in Figure 1,
we have assumed that $\sin^2 2\theta_0^{\mu \mu'} \simeq 1$ and 
$|\delta m^2_{\mu \mu'}| = 10^{-2}\ eV^2$.
Also shown (dashed line) is the cosmology bound
$|\delta m^2_{\tau \mu'}| \stackrel{<}{\sim} 1600 \ eV^2$.
The dashed-dotted line is the nucleosynthesis bound Eq.(\ref{blob2})
assuming that $\delta N_{eff} = \rho'/\rho_{\nu_{\alpha}} 
\stackrel{<}{\sim} 0.6$.
\vskip 0.5cm
\noindent
{\large \bf Acknowledgements}
\vskip 0.5cm
\noindent
This work was supported by the Australian Research Council.


\begin{thebibliography}{99}
\bibitem{bzm}
This point has also been observed in the context of a
mirror Universe model by
Z. G. Berezhiani and R. N. Mohapatra, [Phys. Rev. D52, 6607 (1995)].
In their model the parity symmetry is spontaneously
broken which makes the model essentially distinct from
the exact parity symmetric model\cite{flv} where the parity symmetry is
assumed to be {\it unbroken} by the vacuum.
\bibitem{qqq}
Note that in terms of $a^p, b^p$, the matter
mixing angle $\sin^2 2\theta_m$
and the function $\langle \sin^2 {\tau \over 2L_m} \rangle$
(the brackets $\langle ... \rangle$ denote the average
over collision times, $\tau$) are given by\cite{fv2},
$$\sin^2 2\theta_m = {s^2 \over s^2 + (c - b^p + a^p)^2}$$
$$\langle \sin^2 {\tau \over 2L_m} \rangle =
{1 \over 2}{s^2 + (c - b^p + a^p)^2 \over
x^p + (c^p - b^p + a^p)^2}.$$
For anti-neutrino oscillations, the corresponding quantities
can be denoted by $\sin^2 2\bar \theta_m$
and $\langle \sin^2 {\tau \over 2\bar L_m} \rangle$.
These quantities can be obtained from the above equations by
replacing $a^p \to -a^p$.
\bibitem{hhh}
In deriving the equation for $\Delta$, we have used
$$dn^-_{\nu_{\alpha}} = n_{\gamma}L_{\nu_{\alpha}}{6 \over \pi^2 T^3}
{p^2 e^{p/T} dp \over (1 + e^{p/T})^2} + {\cal O}(L^2_{\nu_{\alpha}}),$$
which can easily be obtained by expanding $dn^-_{\nu_{\alpha}}$
in a power series assuming that the chemical potential is small
(see e.g. \cite{fv2}).
Note that $dn^-_{\nu_{\beta}'}$ can be obtained by replacing $T
\to T'$ in the above equation.
\bibitem{remark}
In our numerical work we found that Eq.(\ref{y3}) 
was approximately valid for
the allowed region with $\sin^2 2\theta_0^{\tau \mu'} \stackrel{>}{\sim}
10^{-6}$. For $\sin^2 2\theta_0^{\tau \mu'} \stackrel{<}{\sim}
10^{-6}$, we found that there were few enough mirror neutrinos
produced so that it did not matter significantly whether the rate of
change of lepton number was obtained from
Eq.(\ref{longeq2},\ref{y7}) or 
Eqs.(\ref{eeqq4},\ref{yq}) (with $L^{(\alpha)} \to L^{\alpha \beta'}$).
\bibitem{remark2}
Note that Eq.(\ref{eeqq3}) [and hence Eq.(\ref{longeq2})] is only 
valid provided that the evolution of the lepton number is 
sufficiently smooth (see Ref.\cite{fv2} for a detailed discussion 
about the region of validity of the static approximation).
In Ref.\cite{fv2} we showed that a necessary condition
for Eq.(\ref{eeqq3}) to be valid is that
$$|{\partial L_{\nu_{\alpha}} \over \partial T}|
\stackrel{<}{\sim} 4 \times 10^{-11} \left({T \over {\rm MeV}}
\right)^4 {1 \over {\rm MeV}}.$$
In our numerical work we found that the above equation
was approximately valid in the allowed region of Figures 1,2
provided that $\sin^2 2\theta_0 \stackrel{<}{\sim} 10^{-5}$.
It would be a useful exercise to check our analysis
(especially in the region where $\sin^2 2\theta_0 
\stackrel{>}{\sim} 10^{-5}$)  by
performing a more accurate study using
the density matrix equations (including the neutrino momentum
distribution).
\bibitem{fn39}
If neutrinos are approximately stable (which would be
expected unless some new interactions exist) then
there is a stringent cosmology bound on the mass
of the neutrinos which implies that $m_{\nu_{\tau}} \stackrel{<}{\sim}
40 \ eV$ (see e.g. the review by G. Gelmini and E. Roulet, in
Ref.\cite{msw}).
Note that $m_{\nu_{\tau}} \stackrel{<}{\sim} 40 \ eV$ implies
that $|\delta m^2_{\tau \beta'}| \stackrel{<}{\sim} 1600 \ eV^2$.
\bibitem{nomad}
The NOMAD and CHORUS experiments are $\nu_{\mu} \to \nu_{\tau}$ 
appearance experiments.  These experiments can potentially measure 
$|\delta m^2_{\tau \mu}|$.  Note that within the context of the
exact parity symmetric model we expect that $|\delta m^2_{\tau \mu}| 
\simeq |\delta m^2_{\tau \mu'}|$. To see this, observe that 
$\delta m^2_{\tau \mu}$ is related 
to $\delta m^2_{\tau \mu'},\ \delta m^2_{\mu \mu'}$
by the relation, $\delta m^2_{\tau \mu} = \delta m^2_{\tau \mu'} -
\delta m^2_{\mu \mu'}$. Thus, for large $|\delta m^2_{\tau \mu}| 
\stackrel{>}{\sim} 1 \ eV^2$, $\delta m^2_{\tau \mu} \simeq
\delta m^2_{\tau \mu'}$ assuming that $|\delta m^2_{\mu \mu'}| 
\stackrel{<}{\sim} 10^{-1} \ eV^2$ (as suggested by the atmospheric
neutrino anomaly).
\end{thebibliography}
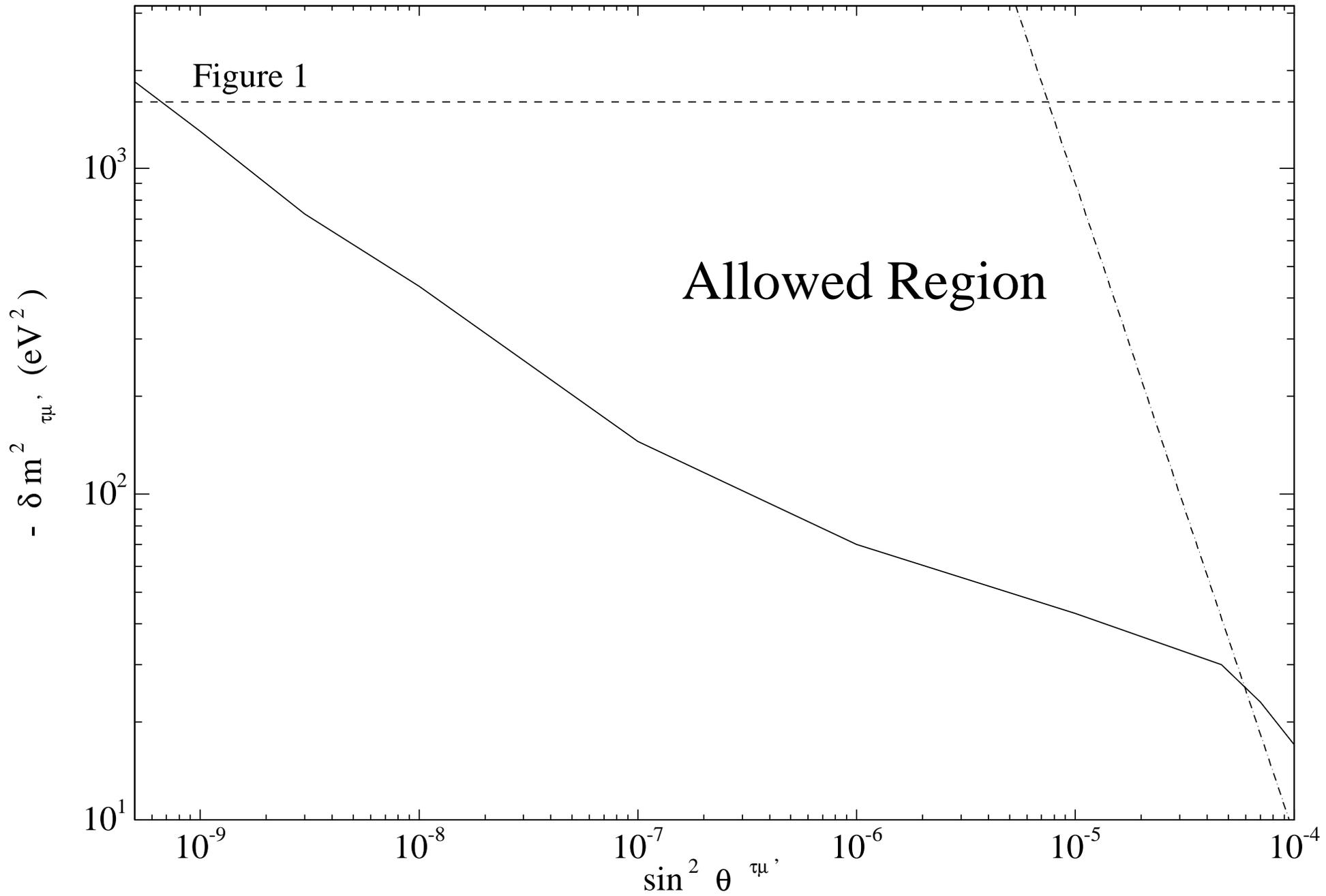
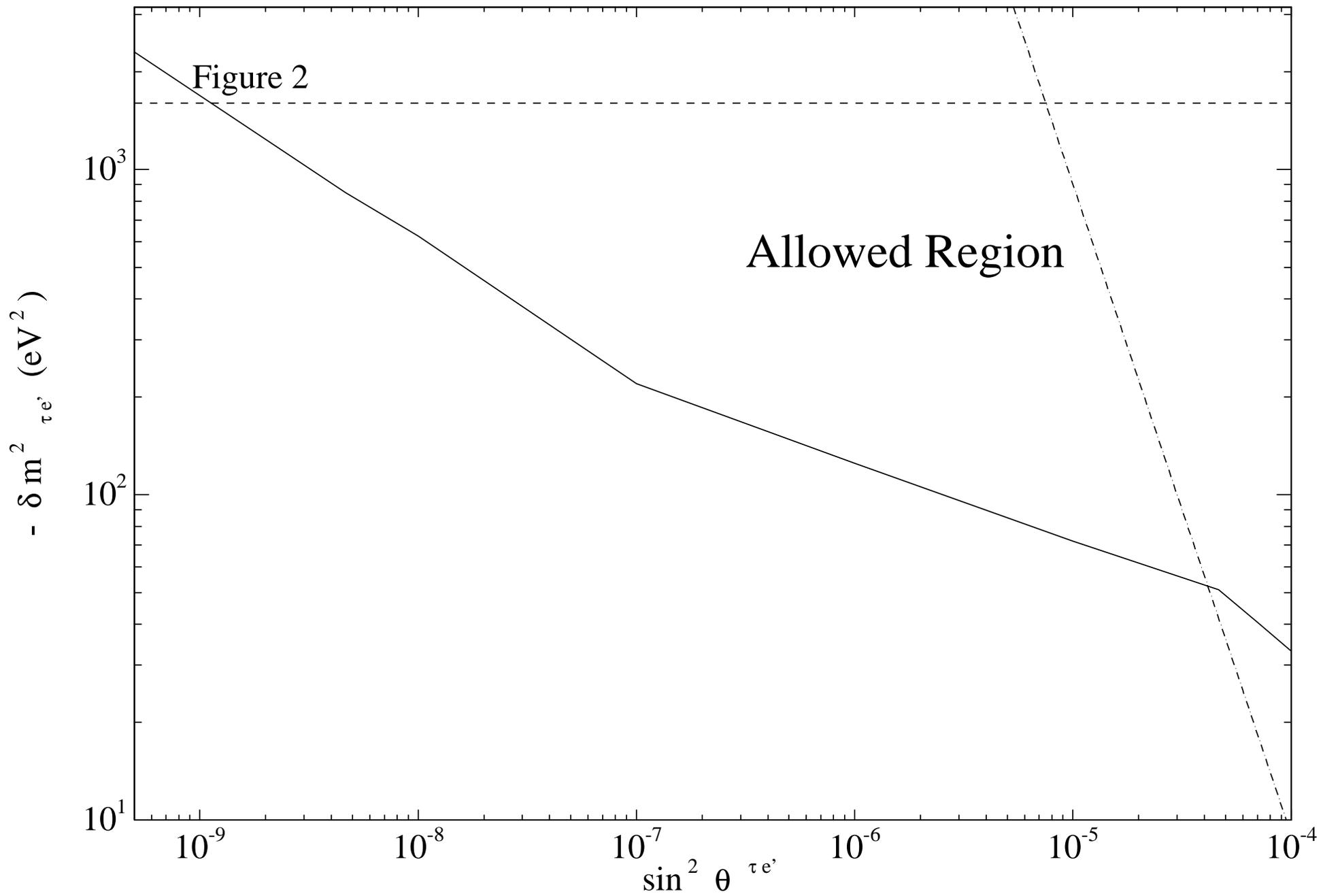
\end{document}